\documentclass[preprint,showpacs,aps,pra]{revtex4-1}
\usepackage{epsfig}
\usepackage{psfrag}
\usepackage{graphicx}
\usepackage{epstopdf}
\usepackage{graphics}
\begin{document}

%\begin{frontmatter}

\title{Neutrino magnetohydrodynamics}

\author{Fernando Haas and Kellen Alves Pascoal}
\affiliation{Instituto de F\'{\i}sica, Universidade Federal do Rio Grande do Sul, Av. Bento Gon\c{c}alves 9500, 91501-970 Porto Alegre, RS, Brasil}

\author{Jos\'e Tito Mendon\c{c}a}
\affiliation{IPFN, Instituto Superior T\'ecnico, Universidade de Lisboa, 1049-001 Lisboa, Portugal}
\affiliation{Instituto de F\'isica, Universidade de S\~ao Paulo, 05508-090 S\~ao Paulo, SP, Brasil}

\begin{abstract}
A new neutrino magnetohydrodynamics (NMHD) model is formulated, 
where the effects of the charged weak current on the electron-ion magnetohydrodynamic
fluid are taken into account. The  model incorporates in a systematic
way the role of the Fermi neutrino weak force in magnetized plasmas. A fast
neutrino-driven short wavelengths  instability associated with the magnetosonic wave is derived. Such
an instability should play a central role in strongly magnetized plasma as occurs
in supernovae, where dense neutrino beams also exist. In addition, in the case of nonlinear or high frequency waves, the neutrino coupling is shown to be responsible for
breaking the frozen-in magnetic field lines condition even in infinite conductivity plasmas. Simplified and ideal NMHD assumptions were adopted and analyzed in detail.
\end{abstract}

%\begin{keyword}
%Neutrino interactions \sep magnetohydrodynamic waves \sep magne\-to\-hy\-dro\-dy\-na\-mics and plasmas \sep neutrino-driven instabilities
%\PACS 13.15.+g \sep 52.35.Bj \sep 95.30.Qd
%\end{keyword}

%\end{frontmatter}
\pacs{13.15.+g, 52.35.Bj, 95.30.Qd}
%\linenumbers

\maketitle

\section{Introduction}

Neutrinos are elusive particles weakly interacting with matter but playing a central role in several still unsolved astrophysical phenomena, including supernova explosions, the formation of structure in the Universe and neutron star core cooling ~\cite{Raffelt}. A significant amount of energy transfer between neutrino beams and plasma waves can take place over distances, thus suggesting that such a mechanism could be crucial for the formation of an outgoing stalled shock in type II supernovae ~\cite{Bingham}. Therefore, collective plasma effects tend to be more crucial than single particle processes, regarding the coupling to neutrinos. Such a coupling is described by the emergence of an effective neutrino charge in an ionized medium ~\cite{Oraevsky, Nieves, Mendo}, producing kinetic and reactive instabilities as well as neutrino Landau damping of plasma waves ~\cite{Silva}.

In the present work a new field of research is proposed, where one of the most popular approaches to space and laboratory plasmas, the magnetohydrodynamic (MHD) theory, is extended in order to incorporate neutrino dynamics. Therefore the contribution bridges the language gap between two major communities, namely astrophysical plasma and particle physicists. In addition, the inclusion of neutrinos should be considered as a new avenue in the study of astrophysical phenomena using laser-produced plasma, in the context of the so-called magneto-quantum-radiative hydrodynamic equations ~\cite{Cross}. 

Naturally the hydrodynamic modeling of neutrino based astrophysical problems is not completely new and has been considered in the past, as in the case of neutrino-driven convection in core-collapse supernova ~\cite{Burrows}. Typically in the previous approaches, neutrinos appear by means of an approximate input of heating and cooling with local prescriptions, acting as a source in the energy transport equation for a neutral fluid. The collective plasma effects are therefore ignored. In particular, the role of the ambient magnetic field is not usually taken into account in a systematic way (see ~\cite{Murphy} for a review). One has therefore a language dichotomy from neutrino particle and plasma physics communities. An intermediate setting containing the essential aspects from neutrino theory and collective plasma aspects in a sufficiently simple MHD description would be a welcome tool to fill the referred language gap, stimulating advances in the field. 

Recently, neutrino-plasma fluid models have been proposed, first in a purely electrostatic context ~\cite{Serbeto} and then ~\cite{Mendonca} allowing for magnetic fields and neutrino flavor oscillations ~\cite{Duan}. In the following, the discussion of neutrino-based magnetic field structures is systematized in terms of a modified MHD theory, to be called neutrino magnetohydrodynamics (NMHD). The derivation is based on a two-fluid plasma model coupled to a neutrino species, taking into account the charged weak current. In view of the complexity of the resulting system of equations, standard assumptions toward the simplified and ideal MHD theory ~\cite{Akhiezer, Bellan, Bittencourt} will be adopted. In spite of the overall simplicity, the neutrinos will be shown to be responsible for qualitatively new phenomena, such as magnetic field lines diffusion (in a formally infinite conductivity plasma) and a fast new beam instability in a magnetosonic waves configuration. Such an instability should play a central role in strongly magnetized plasma as occurs in supernovae. Electrons and ions will be taken as non-relativistic, together with (ultra-)relativistic neutrinos. 

\section{Basic model}

We start with the two-fluid equations for an electron-ion plasma coupled to a neutrino species, following the model put forward in 
~\cite{Mendonca}. The mass and momentum transport equations for electrons (with mass $m_e$ and charge $-e$) are resp. 
\begin{eqnarray}  
 \frac{\partial n_e}{\partial t} + \nabla \cdotp (n_e \textbf{u}_e) &=& 0 \,, \label{eq01} \\
 m_e \left(\frac{\partial \textbf{u}_e}{\partial t} + \textbf{u}_e \cdotp \nabla \textbf{u}_e \right) = &-& \frac{\nabla P_e}{n_e} - e (\textbf{E} + \textbf{u}_e \times \textbf{B}) \nonumber \\ &+& \textbf{F}_\nu  + {\bf K}_{ei} \,,  \label{eq02}
\end{eqnarray}
while ions (with mass $m_i$ and charge $e$) satisfy 
\begin{eqnarray}
 \frac{\partial n_i}{\partial t} + \nabla \cdotp (n_i \textbf{u}_i) &=& 0 \,, \label{eq03} \\ 
 m_i \left(\frac{\partial \textbf{u}_i}{\partial t} + \textbf{u}_i \cdotp \nabla \textbf{u}_i \right) = &-& \frac{\nabla P_i}{n_i} + e (\textbf{E} + \textbf{u}_i \times \textbf{B})  + {\bf K}_{ie} \,. % &-& \nu_{ie}(\textbf{u}_i -\textbf{u}_e) \,. 
\label{eq04}
\end{eqnarray}
Finally, the neutrino fluid satisfy 
\begin{eqnarray}
 \frac{\partial n_\nu}{\partial t} + \nabla \cdotp (n_\nu \textbf{u}_\nu) &=& 0 \,, \label{eq05} \\ 
 \frac{\partial \textbf{p}_\nu}{\partial t} + \textbf{u}_\nu \cdotp \nabla \textbf{p}_\nu &=& \sqrt{2}\,G_F (\textbf{E}_e + \textbf{u}_\nu \times \textbf{B}_e) \label{eq06} \,.
\end{eqnarray}
Apart from the neutrino component, Eqs. (\ref{eq01})-(\ref{eq06}) are the traditional two-fluid (electron plus ion) plasma equations \cite{Akhiezer, Bellan, Bittencourt}, which are the starting point for the magnetohydrodynamic (by definition, an one-fluid) plasma model. 
In the basic equations,  $n_{e,i,\nu}$ and ${\textbf u}_{e,i,\nu}$ are resp. the electron, ion and neutrino number densities and fluid velocities, and ${\textbf p}_\nu = \mathcal{E}_\nu {\textbf{u}_\nu}/{c^2}$ is the momentum of the relativistic neutrino beam having and energy $\mathcal{E}_\nu$, where $c$ is the speed of light. Moreover, $m_{e,i}$ and $P_{e,i}$ denote the electron-ion masses and fluid pressures and ${\bf E}, {\bf B}$ are the electric and magnetic fields, while the neutrino force $\textbf{F}_\nu$ is
\begin{equation}
\textbf{F}_\nu = \sqrt{2}\,G_F (\textbf{E}_\nu + \textbf{u}_e \times \textbf{B}_\nu) \,, \label{eq07}
\end{equation}
where $G_F$ the Fermi constant of weak interaction and $\textbf{E}_\nu$, $\textbf{B}_\nu$ are effective fields induced by the weak interaction, 
\begin{eqnarray}
 \textbf{E}_\nu = - \nabla n_\nu - \frac{1}{c^2}\frac{\partial }{\partial t}(n_\nu\textbf{u}_\nu) \,, \label{eq08} \quad 
 \textbf{B}_\nu = \frac{1}{c^2}\nabla \times (n_\nu\textbf{u}_\nu)  \,, \label{eq09}
\end{eqnarray} 
jointly with 
\begin{equation}
 \textbf{E}_e = - \nabla n_e - \frac{1}{c^2}\frac{\partial }{\partial t}(n_e\textbf{u}_e) \,,  \quad 
 \textbf{B}_e = \frac{1}{c^2}\nabla \times (n_e\textbf{u}_e)  \,, \label{eq11}
 \end{equation} 
to be inserted in Eq. (\ref{eq06}). Note that only the charged weak current was retained, disregarding the neutral weak current which would lead to a correction of order one to the terms proportional to $G_F$. This is because electrons are coupled to electron neutrinos by the charged bosons $W^{\pm}$, while both protons and electrons are coupled to all neutrino flavors by the neutral boson $Z$. The weak interactions between neutrinos and background electrons is associated to neutrino angular momentum in a plasma vortex ~\cite{vortex} and to electrostatic instabilities in fully degenerate plasmas ~\cite{Rios}. A more detailed discussion of the neutrino-plasma coupling is given in Appendix A, for completeness.

The modifications in comparison with the model in ~\cite{Mendonca} are the inclusion of mobile ions and of a momentum transfer between electron and ion fluids, as follows from the straightforward derivation of fluid equations from for two-species kinetic theory \cite{Akhiezer, Bellan, Bittencourt}, described by the terms ${\bf K}_{ei, ie}$  which are resp. the rates of change in the electron (ion) fluid momentum due to collisions with ions (electrons). Notice there's no electron-electron or ion-ion collision terms because the electron (ion) fluid can not cause a drag to itself. On the same footing, there can be no drag of the entire (electron plus ion) MHD fluid, so that, by definition, ${\bf K}_{ei} + {\bf K}_{ie} = 0$. Therefore, the specific form of the dissipation terms is irrelevant as far as total momentum conservation is assured, as discussed at length in the common place derivation of MHD theory ~\cite{Akhiezer, Bellan, Bittencourt}. Nevertheless, it is useful to adopt the usual phenomenological expressions
\begin{equation}
{\bf K}_{ei} = - m_e \nu_{ei} ({\bf u}_e - {\bf u}_i) \,, \quad {\bf K}_{ie} = - m_i \nu_{ie} ({\bf u}_i - {\bf u}_e) \,, 
\end{equation}
which are the first order Taylor expansions of the drag terms in powers of the electron and ion velocities difference, in terms of the collision frequency coefficients 
$\nu_{ei}$ and $\nu_{ie}$. Global momentum conservation in collisions imply $m_e\nu_{ei}=m_i\nu_{ie}$, so that $\nu_{ei}\ll \nu_{ie}$ since $m_i \gg m_e$. The specific form of the dissipation terms is irrelevant as far as total momentum conservation is assured, as discussed at length in the common place derivation of MHD theory ~\cite{Bittencourt, Bellan}. Moreover, for simplicity neutrino flavor oscillations are presently disregarded. 

Closure is provided by Maxwell's equations,  
\begin{eqnarray}
 \nabla \cdot \textbf{E} &=& \frac{\rho}{\varepsilon_0} \,, \quad 
 \nabla \cdot \textbf{B} = 0 \,, \label{eq13} \nonumber \\
 \nabla \times \textbf{E} &=& - \frac{\partial \textbf{B}}{\partial t} \,, \quad 
 \nabla \times \textbf{B} = \mu_0 \textbf{J} + \frac{1}{c^2}\frac{\partial \textbf{E}}{\partial t} \,. \label{eq15}
\end{eqnarray}
where $\varepsilon_0$ and $\mu_0$ are the vacuum permittivity and permeability and the charge and current densities are given respectively by
\begin{equation} 
 \rho = e (n_i - n_e) \,, \quad %\label{eq16} \\
 \textbf{J} = e (n_i \textbf{u}_i - n_e \textbf{u}_e)\,.  \label{eq17} 
\end{equation}

Eqs.~(\ref{eq01})-(\ref{eq15}) constitute a complete neutrino-plasma interaction hydrodynamic model allowing to obtain, among many possibilities, a magnetohydrodynamic formulation where electron and ion fluids are mixed. For this purpose, we introduce the global mass density $\rho_m$ and the global fluid velocity $\textbf{U}$, 
\begin{equation}
 \rho_m = m_en_e + m_in_i \,, \quad \textbf{U} = \frac{m_en_e\textbf{u}_e + m_in_i\textbf{u}_i}{m_en_e + m_in_i} \,. \label{eq20} 
\end{equation}

Following the standard procedure, taking into account $m_i \gg m_e$ whenever possible, we obtain the mass and momentum transport equations,
\begin{eqnarray}
 \frac{\partial \rho_m}{\partial t} + \nabla \cdot (\rho_m \textbf{U}) &=& 0 \,,  \label{eq26} \\
 \rho_m \left(\frac{\partial \textbf{U}}{\partial t} + \textbf{U} \cdot \nabla \textbf{U} \right) = &-& \nabla \cdot \Pi + \rho \textbf{E} + \textbf{J} \times \textbf{B} \nonumber \\ &+&  \left(\frac{\rho_m}{m_i} - \frac{\rho}{e}\right) \textbf{F}_\nu \,. \label{eq27}
\end{eqnarray}
with the pressure dyad
\begin{equation} 
\label{ppp}
\Pi = P \,\textbf{I} + \frac{m_em_in_en_i}{\rho_m} (\textbf{u}_e - \textbf{u}_i) \otimes (\textbf{u}_e - \textbf{u}_i) \,,
\end{equation}
where $P = P_e + P_i$ is the total plasma scalar pressure, $\textbf{I}$ is the identity matrix and $\otimes$ denotes the tensor product. Following the standard treatment ~\cite{Bittencourt}, the second term on the right-hand side of Eq. (\ref{ppp}) will be disregarded in view of scalar pressure dominated conditions.

Taking the time-derivative of Amp\`ere-Maxwell's law and using the same procedure of standard MHD ~\cite{Akhiezer, Bellan, Bittencourt}, a generalized Ohm's law can be derived, 
\begin{eqnarray}
  \frac{m_e m_i}{\rho_m e} \frac{\partial \textbf{J}}{\partial t} &-& \frac{m_i}{\rho_m}\nabla P  = e(\textbf{E} + \textbf{U} \times \textbf{B}) \nonumber \\ &-& \frac{m_i}{\rho_m}\textbf{J} \times \textbf{B} - \textbf{F}_\nu -\frac{\textbf{J}}{\sigma} \,, \label{eq30}
\end{eqnarray}
where $\sigma = \rho_m e^2/(m_e m_i \nu_{ei})$ is the longitudinal electric conductivity.

In Eqs. (\ref{eq27}) and (\ref{eq30}) one has the neutrino force (\ref{eq07}), which is re-expressed as
\begin{equation}
 \textbf{F}_\nu = \sqrt{2}\,G_F \left[\textbf{E}_\nu + \left(\textbf{U} - \frac{m_i \textbf{J}}{\rho_m e}\right) \times \textbf{B}_\nu \right] \,, \label{eq29}
\end{equation}
representing the net neutrino influence on the MHD fluid. The plasma back-reacts on the neutrino fluid through the effective fields ${\textbf E}_e, {\textbf B}_e$ defined in Eq. (\ref{eq11}), which are the source fields in the neutrino moment equation (\ref{eq06}). 

In view of the extension of the resulting model,  extra assumptions should be adopted, in accordance with the usual procedure but keeping the salient modifications due to the neutrino beam. Under the simplified and ideal MHD conditions ~\cite{Bittencourt}, it will be assumed: (a) formally infinite conductivity $\sigma \to \infty$, so that local charge unbalance can be disregarded, or $\rho \approx 0, n_e \approx n_i$; (b) neglect of the time-derivative of the current density and of the pressure term  in Eq. (\ref{eq30}) in view resp. of slow time dependence and magnetic dominated (low-beta) plasma situation; (c) in the same Eq. (\ref{eq30}) we neglect the Hall term $\sim \textbf{J}\times\textbf{B}$ in view of a high collision frequency in comparison to the gyro-frequency. Keeping this contribution would correspond to a more complex Hall NMHD ~\cite{Bellan}, which could in principle give rise to interesting phenomena to be analyzed in the future; (d) disregard relativistic corrections on MHD equations, since electrons and ions are assumed non-relativistic. In the same spirit, for waves with phase velocity much smaller than the speed of light, the displacement current can be neglected in the Amp\`ere-Maxwell law; (e) adoption of the equation of state $\nabla P = V_{S}^2 \nabla \rho_m$, where $V_S$ is the adiabatic speed of sound. 

The above standard assumptions (a)-(e) allow to eli\-mi\-na\-te the electric field which becomes 
\begin{equation}
\label{ee}
\textbf{E} = - \textbf{U}\times\textbf{B} + \textbf{F}_{\nu}/e \,,
\end{equation}
containing a neutrino force correction. Moreover, in a non-relativistic electron-ion fluid the effective fields in (\ref{eq11}) simplify to
\begin{equation}
\textbf{E}_e = - \nabla n_e = - \nabla\rho_m/m_i \,, \quad \textbf{B}_e = 0 \,,
\end{equation}
where quasi-neutrality was also used. 

We are now in a position to enumerate the basic equations of the simplified and ideal NMHD model. They are: 
(i) the neutrino continuity equation (\ref{eq05}); (ii) the neutrino force equation (\ref{eq06}), re-expressed as 
\begin{equation}
\frac{\partial \textbf{p}_\nu}{\partial t} + \textbf{u}_\nu \cdotp \nabla \textbf{p}_\nu = - \frac{\sqrt{2}\,G_F}{m_i} \nabla \rho_m \,. \label{nf}
\end{equation}
(iii) the MHD continuity equation (\ref{eq26}); (iv) the MHD force equation (\ref{eq27}), re-expressed as 
\begin{equation}
\frac{\partial \textbf{U}}{\partial t} + \textbf{U} \cdot \nabla \textbf{U} = - \frac{V_{S}^2 \nabla \rho_m}{\rho_m} + \frac{(\nabla\times\textbf{B}) \times \textbf{B}}{\mu_0 \,\rho_m} + \frac{\textbf{F}_\nu}{m_i} \,. \label{eq34}
\end{equation}
(v) Faraday's law which reads
\begin{equation}
\frac{\partial\textbf{B}}{\partial t} = \nabla\times\left(\textbf{U}\times\textbf{B} - \frac{\textbf{F}_{\nu}}{e}\right) \,, \label{eq37}
\end{equation}
after eliminating the electric field, and considering the magnetic Gauss's law as initial condition.
In the model equations, the neutrino force $\textbf{F}_\nu$ is defined in Eq. (\ref{eq29}) where $\textbf{J} = \nabla\times\textbf{B}/\mu_0$, containing the effective fields $\textbf{E}_\nu, \textbf{B}_\nu$ found from Eq. (\ref{eq09}). In this way we have a complete set of 11 equations for 11 variables, namely $\rho_m, n_\nu$ and the components of $\textbf{U}, \textbf{p}_\nu$ and $\textbf{B}$.

An immediate possible consequence of the neutrino coupling is that frozen-in magnetic field lines can no longer exist, in view of the neutrino force in Eq. (\ref{eq37}). This qualitatively new effect comes from the weak force acting on the electrons, and hence on the MHD fluid, which is the source of the magnetic field itself. However, in quasi-static situations where $\textbf{U} \approx 0, \textbf{J} \approx 0$ and near equilibrium, the term containing the neutrino ``weak'' magnetic field $\textbf{B}_\nu$ in Eq. (\ref{eq29}) is of second-order. Moreover, for subluminal and low-frequency waves the weak force reduces to $\textbf{F}_\nu = - \sqrt{2}\,G_F \nabla n_\nu$, so that $\nabla\times\textbf{F}_\nu = 0$ and the frozen-in condition is still satisfied as seen from Eq. (\ref{eq37}). More general, nonlinear and/or high frequency neutrino perturbations can produce magnetic field lines diffusion, even in a simplified and ideal MHD model. 

\section{Linear waves and instabilities}

It is important to assure the validity conditions of the simplified and ideal NMHD equations. Since neutrinos are almost always a perturbation, to zeroth order these validity conditions are the same as for ideal MHD, which are described e.g. in Ref. \cite{Akhiezer}. Starting with the two-fluid (electron and ion) species plasma model, a possible justification for a MHD model (by definition, always an one-fluid model) is provided by a high collisional rate, or 
\begin{equation}
\label{sf}
|\omega| \ll \nu_{ie} \,, %\approx \nu_{ee} \,,
\end{equation}
where $|\omega|^{-1}$ is the time-scale of changes of the MHD flow, $\nu_{ie}^{-1}$ is the time-scale of the ion fluid momentum changes due to collision against electrons. % and $\nu_{ee}$ is the electron-electron collision frequency. 
In addition, the simplified and ideal MHD equations are valid for high conductivity plasma and a typical MHD speed $V << c$, or (as shown in Ref. \cite{Akhiezer}, Eqs. (1.5.2.8)) 
\begin{equation}
\label{id}
\frac{\varepsilon_0\,|\omega|}{\sigma} \ll 1 \,, \quad \frac{\varepsilon_0 \,V}{\sigma\,L} \ll 1 \,,
\end{equation}
where $L$ is a characteristic length scale. In astrophysical settings only the first in (\ref{id}) can pose difficulties. In view of the expression of the conductivity below Eq. (\ref{eq30}), the combination of Eqs. (\ref{sf}) and (\ref{id}) expressed in terms of $\nu_{ei}$ is 
\begin{equation}
\label{fk}
\frac{m_i \,|\omega|}{m_e} \ll \nu_{ei} \ll \frac{\omega_{pe}^2}{|\omega|} \,,
\end{equation}
where for an equilibrium number density $n_0$ one has $\rho_m \approx n_0 m_i$ and where $\omega_{pe} = \sqrt{n_0 e^2/(m_e \varepsilon_0)}$. To summarize, the first inequality in Eq. (\ref{fk}) assures the description of a single conducting fluid; the second inequality assures ideality so that there is no wave damping in this framework (also viscous effects are disregarded). Nevertheless, it should be keep in mind that by definition kinetic effects such as electron and neutrino Landau damping are not included in an hydrodynamic model. 

Equation (\ref{fk}) can be expressed in terms of more specific physical parameters using (Ref. \cite{Spitzer}, chapter V) the Landau electron-electron collision frequency 
\begin{equation}
\label{spit}
\nu_{ee} \approx \nu_{ei} = \frac{2\,\omega_{pe}}{3}\frac{\ln\Lambda}{\Lambda} \,, \quad \Lambda = 4\pi n_0\lambda_{D}^3/3 \,, \quad \lambda_D = v_T/\omega_{pe} \,, 
\end{equation}
where $v_T = \sqrt{2\kappa_B T_e/m_e}$ is the thermal speed for an electron fluid temperature $T_e$ and $\kappa_B$ is the Boltzmann constant. These expressions apply for slight degeneracy and relativistic effects for electrons. Implicitly, a weak coupling condition $\Omega \gg 1$ is also assumed (equivalently, $\nu_{ei} \ll \omega_{pe}$). Then from Eqs. (\ref{fk}) and (\ref{spit}) we get in a dimensionless form 
\begin{equation}
\label{thefck}
\frac{m_i}{m_e}\frac{|\omega|}{\omega_{pe}} \ll \frac{2}{3}\frac{\ln\Lambda}{\Lambda} \ll \frac{\omega_{pe}}{|\omega|} \,.
\end{equation}

Alternative closure schemes not based on collisional estimates but e.g. on a high magnetic field assumption \cite{Braginskii} will be not addressed here, for simplicity. A more detailed discussion of the validity conditions for ideal magnetohydrodynamics can be found e.g. in 
Ref. \cite{Balescu}, chapter VII.

As an illustration of decisive consequences of the neutrino coupling, small amplitude perturbations around an homogeneous magnetized equilibrium $\rho_m = \rho_{m0}, \, n_{\nu} = n_{\nu 0}, \, \textbf{U} = 0, \, \textbf{p}_\nu = \textbf{p}_{\nu 0}, \, \textbf{B} = \textbf{B}_0$ will be analyzed. Linea\-ri\-zing Eqs. (\ref{eq05}), (\ref{eq26}) and (\ref{nf})-(\ref{eq37}) considering plane waves of frequency $\omega$, wave-vector $\textbf{k}$ the result is 
\begin{eqnarray}
 \omega^2\delta\textbf{U} &=& \left(V^2_S\!+ V^2_A +\! V^2_N \frac{(c^2k^2 - (\textbf{k}\cdot\textbf{u}_{\nu 0})^2)}{(\omega- \textbf{k}\cdot \textbf{u}_{\nu 0})^2}\right)\!(\textbf{k}\cdot\delta\textbf{U})\textbf{k} \nonumber \\ &+& (\textbf{k} \cdot \textbf{V}_A)\Bigl((\textbf{k} \cdot \textbf{V}_A)\delta\textbf{U} - (\delta\textbf{U}\cdot\textbf{V}_A)\textbf{k} \nonumber \\ &-& (\textbf{k}\cdot\delta\textbf{U})\textbf{V}_A\Bigr)  \,, \label{eq51}
\end{eqnarray}
after eliminating all variables except the MHD fluid velocity perturbation $\delta\textbf{U}$.  Whenever harmless, the low frequency assumption $\omega/k \ll c$ was used. In Eq. (\ref{eq51}), the vector Alfv\'en velocity $\textbf{V}_A$ and a new characteristic ``neutrino speed" $V_N$ were employed. These are given by                  
\begin{eqnarray}
\textbf{V}_A = \frac{\textbf{B}_0}{(\rho_{m0} \mu_0)^{1/2}} \,, \quad 
V_N = \left(\frac{2G^2_F \rho_{m0} n_{\nu0}}{m^2_i \mathcal{E}_{\nu 0}}\right)^{1/2} \,, \label{eq53}
\end{eqnarray}
where $\mathcal{E}_{\nu 0}$ is the equilibrium neutrino beam energy so that $\textbf{p}_{\nu 0} = \mathcal{E}_{\nu 0} \textbf{u}_{\nu 0}/c^2$. It is interesting to note that $V_N$ is determined by both MHD and neutrino variables, emphasizing the coupling between them. Equation (\ref{eq51}) is the standard general MHD dispersion relation (as shown in Eq. (2.21) in Ref. ~\cite{Bittencourt}), except for the neutrino contribution. 

\begin{figure}[h]
\begin{center}
\includegraphics[width=2.5in]{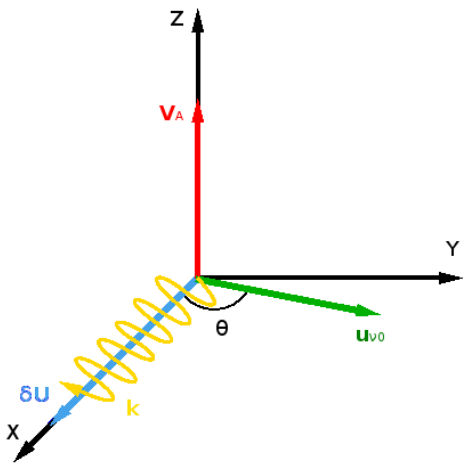}
% \onefigure[angle=0,scale=0.47]{fig1.eps}
% \includegraphics[angle=0,scale=0.45]{fig1.eps}
\caption{Geometry for the instability shown in Eq. (\ref{ins}).}
\label{fig1} 
\end{center}
\end{figure}

From inspection of the dispersion relation (\ref{eq51}), it is seen that purely transverse waves with $\textbf{k} \perp \textbf{V}_A$ and $\textbf{k} \perp \delta\textbf{U}$ are not affected by the neutrino beam. Hence Alfv\'en waves are not perturbed, at least under the present set of approximations. If instead we consider the important case of magnetosonic (fast Alfv\'en) waves with $\textbf{k} \perp \textbf{V}_A$ and $\textbf{k} \parallel \delta\textbf{U}$ a neutrino-driven instability is found. For generality, an angle $\theta$ between the wave propagation and the neutrino beam can be allowed, as shown in Fig. \ref{fig1}, so that $\textbf{k} \cdot \textbf{u}_{\nu 0} = k\, u_{\nu 0} 
\cos\theta$. The dispersion relation then reduces to 
\begin{equation}
 \left(\frac{\omega}{k} - u_{\nu0} \cos\theta\right)^2 \left(\frac{\omega^2}{k^2}-V^2_S-V_A^2\right) = V_N^2 (c^2-u^2_{\nu0}\cos^2\theta) \,. \label{n}
\end{equation}
The right-hand side of Eq. (\ref{n}) can be taken as a perturbation. Therefore, focusing on the unstable mode we consider the neutrino-beam mode $\omega = k\,u_{\nu 0}\cos\theta + i\gamma$, where $\gamma$ is much smaller than the magnetosonic frequency $\Omega \equiv (V_{S}^2 + V_{A}^2)^{1/2} k$. The approximate solution is 
\begin{equation}
 \gamma = \frac{V_N k (c^2 - u^2_{\nu0}\cos^2 \theta)^{1/2}}{(V^2_S+V^2_A-u^2_{\nu0}\cos^2\theta)^{1/2}} \,, \label{ins}
\end{equation}
pointing for an instability ($\gamma > 0$) provided  $V^2_S+V^2_A > u_{\nu0}^2 \cos^2\theta$. In view of the ultra-relativistic neutrinos ($u_{\nu 0} \approx c$), the instability is more likely for perpendicular propagation, $\theta = \pi/2$. In this case, the ultra-relativistic neutrino beam velocity appears only implicitly, by means of the neutrino beam energy $\mathcal{E}_{\nu 0}$ contained in $V_N$. Specific features were identified, namely: the instability is larger for $\textbf{u}_{\nu 0} \perp \textbf{k}$ and is suppressed for pa\-rallel pro\-pa\-ga\-tion; as expected, the instability is larger for denser neutrino beam and smaller ambient 
magnetic field. In addition, the growth rate turns out to scale as $\gamma \sim V_N \sim G_F$, which is much larger than typical electrostatic neutrino-plasma beam instabilities ~\cite{Bingham, Silva} which have no connection with the ambient magnetic field.

In the case of per\-pen\-di\-cu\-lar neutrino propagation ($\theta = \pi/2$) and neglecting the adiabatic sound speed in comparison to the Alfv\'en speed in a strongly magnetized plasma, the result is
\begin{equation}
\label{gg}
\gamma = \left(\frac{2\,n_{\nu 0}}{\varepsilon_0 \mathcal{E}_{\nu 0}}\right)^{1/2}\frac{G_F n_0 k}{B_0} \,.
\end{equation}
Using the Fermi constant $G_F = 1.45 \times 10^{-62}\, {\rm J.m}^3$, for hydrogen plasma and typical ~\cite{Burrows} supernova pa\-ra\-me\-ters $n_{\nu 0} = 10^{35}\,{\rm m}^{-3}$ which is the same as the MHD fluid number density and  $\mathcal{E}_{\nu 0} = 10\, {\rm MeV}$, per\-pen\-di\-cu\-lar neutrino propagation ($\theta = \pi/2$) and neglecting the adiabatic sound speed in comparison to the Alfv\'en speed in a strongly magnetized plasma, the result is $\gamma = 172.33\, k/B_0, \Omega = 6.90 \times 10^{-2} B_0 k$, where S. I. units are employed. In this case one has $\gamma/\Omega = 2.50 \times 10^{3} B_{0}^{-2} \ll 1$ for the strong magnetic fields $B_0 \approx 10^6 - 10^8 \, {\rm T}$ appearing in core-collapse events. Hence the growth rate is much smaller than the magnetosonic frequency, justifying the approximation used in the derivation of Eq. (\ref{ins}). One might consider magnetic field strengths below the electron Schwinger critical QED field $B_c = m_{e}^2 c^2/(e \hbar) = 4.42 \times 10^{9}\, {\rm T}$, but large enough to discard $V_S \ll V_A$. 

\begin{figure}[h]
\begin{center}
\includegraphics[width=3.5in]{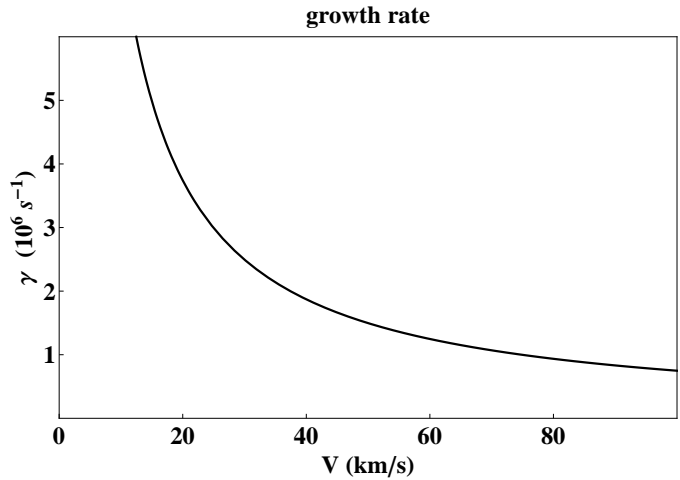}
% \onefigure[angle=0,scale=0.45]{fig2.eps}
%\includegraphics[angle=0,scale=0.45]{fig1.eps}
\caption{Growth rate from Eq. (\ref{n}) as a function of the magnetosonic speed $V = \sqrt{V_{S}^2+V_{A}^2}$. 
Parameters: $\theta = \pi/2, \, V_N = 3.97 \times 10^{-8}\, {\rm m/s}, \, k = 2\pi \times 10^{9} \,{\rm m}^{-1}$.}
\label{fig2}
\end{center}
\end{figure}

For the sake of illustration, one might consider $\kappa_B T_e = 0.1\, {\rm MeV}$, so that $\Lambda = 487.38$, $\omega_{pe} = 1.78 \times 10^{19} \,{\rm s}^{-1}$, $\nu_{ei} = 1.51 \times 10^{17}\,s^{-1}$. The chain of inequalities (\ref{thefck}) becomes, with $|\omega| \approx \gamma$ and for electron-proton plasma for simplicity, 
\begin{equation}
1.77 \times 10^{-14} k/B_0 \ll 8.47 \times 10^{-3} \ll 1.03 \times 10^{17} B_0/k \,,
\end{equation}
which is well attended for any reasonable wavenumber for the strong magnetic fields of interest. Therefore the simplified and ideal MHD conditions are satisfied. Just as an example, one might consider $B_0 = 10^6 \,{\rm T}$ and a wavelength $\lambda = 2\pi/k = 1 \,{\rm nm}$ in the soft X-ray range. Then from Eq. (\ref{gg}) one has $\gamma = 1.08 \times 10^6 \,{\rm s}^{-1}$. This could to be compared to the time-scale (around 1 sec.) of the supernova explosion. Hence the new neutrino-driven instability is fast enough to be an excellent candidate to trigger the cataclysmic event. In addition, $V_A = 69.03 \, {\rm km/s}, V_N = 3.97 \times 10^{-8} \,{\rm m/s}$. For the same parameters set except that the Alfv\'en velocity and adiabatic speed of sound are left free, 
one might calculate the growth rate from the unstable branch of the dispersion relation (\ref{n}) as a function of the magnetosonic speed $V = \sqrt{V_{S}^2 + V_{A}^2}$ as shown in Fig. \ref{fig2}. %In this case the maximum growth rate formally obtained for $V \rightarrow 0$ is $\gamma_{\rm max} = (V_N c)^{1/2} \,\,k = 1.45 \times 10^{13} \,{\rm s}^{-1}$.} 

One might for instance put on question the neglect of the displacement current. However, one has %on basis of Fig. \ref{fig2} taking $\gamma \sim 10^{10} s^{-1}$ one has 
\begin{equation}
\frac{\varepsilon_0 |\partial{\bf E}/\partial t|}{\sigma |{\bf E}|} \sim \frac{\varepsilon_0 \gamma}{\sigma} \sim \frac{\nu_{ei}\,\gamma}{\omega_{pe}^2} \ll 1
\end{equation}
which is automatically satisfied in view of the last inequality in Eq. (\ref{fk}).
%which is still quite small for usually very large conductivities $\sigma$ greater than $1 \,{\rm mho}/{\rm m}$. 
Another concern is about possible mechanisms for the anisotropic neutrino velocities distribution associated to the neutrino beam, which have been discussed elsewhere ~\cite{Laming}. In particular, far from the the neutrinosphere  there is a small angular spread of the radially directed neutrino beam.  Moreover in type II supernovae, the neutrinos are known to be sufficiently collimated to provide a suitable electrostatic instability mechanism ~\cite{Silva2}.

%$\strut$
%\vspace{.3cm}
%$\strut$

The above results are wavenumber-dependent. For more generality one might consider $\theta = \pi/2$ for simplicity, so that $\gamma = V_N c\, k/V$  clearly satisfying the low-frequency assumption $\gamma/(V k) = V_N c/V^2 \ll 1$ except for extremely small magnetosonic speeds. The result shown in Fig. 3 imply a smaller growth rate for longer wavelengths, but still attaining appreciable values for typical parameters. %It is expected that the excited plasma waves release their energy to the plasma environment by collisional damping driving the supernova shock by heating ~\cite{Serbeto}.  

\begin{figure}[h]
\begin{center}
\includegraphics[width=4.5in]{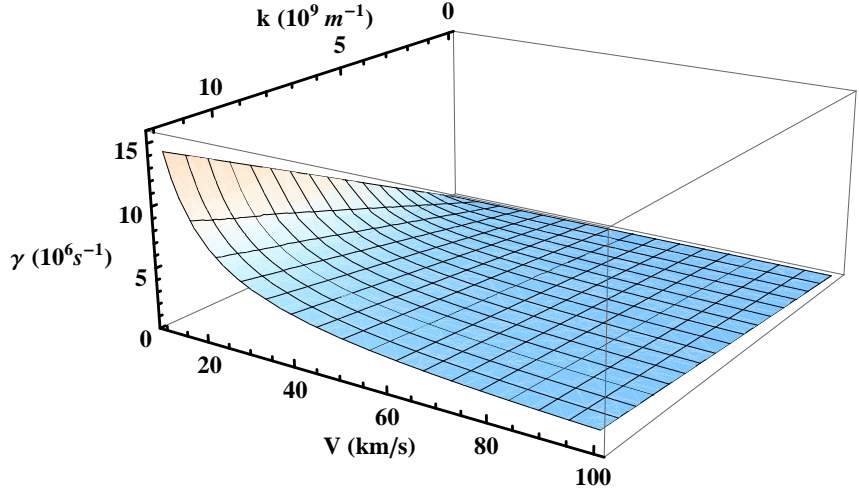}
%\onefigure[angle=0,scale=0.45]{fig3.eps}
%\includegraphics[angle=0,scale=0.45]{fig1.eps}
\caption{Growth rate $\gamma =  V_N c\, k/V$ as a function of the wavenumber $k$ and the magnetosonic speed $V = \sqrt{V_{S}^2+V_{A}^2}$. 
Parameters: $\theta = \pi/2, \, V_N = 3.97 \times 10^{-8}\, {\rm m/s}$.}
\label{fig3} 
\end{center}
\end{figure}

\section{Conclusions}

To summarize, a NMHD model was introduced and analyzed in more detail in the simplified and ideal conditions. The neutrino component was shown to be a sui\-ta\-ble source of magnetic field lines diffusion. In addition, a new neutrino-driven instability was found, associated with the magnetosonic wave geometry. The instability rate FVcan be rather large in core-collapse supernova scenarios, increasing for shorter wavelengths. The full investigation of the dispersion relation (\ref{eq51}) as well as of further ingredients such as finite conductivity, displacement current, Hall NMHD dynamics and nonlinear effects is a fruitful avenue for future research. 

{\bf Acknowledgments}: 
F.~H.~ and J.~T.~M.~ acknowledge the support by Con\-se\-lho Na\-cio\-nal de De\-sen\-vol\-vi\-men\-to Cien\-t\'{\i}\-fi\-co e Tec\-no\-l\'o\-gi\-co (CNPq) and EU-FP7 IRSES Programme (grant 612506 QUANTUM PLASMAS FP7-PEOPLE-2013-IRSES), and K.~A.~P.~ack\-now\-ledges the support by Coordena\c{c}\~ao de Aperfei\c{c}oamento de Pessoal de N\'{\i}vel Superior (CAPES). 

\appendix
\section{Notes on the electron-neutrino interaction}

For completeness and for the convenience of the reader, it is useful to briefly review the key points regarding the electroweak interaction terms in Eqs. (\ref{eq02}) and (\ref{eq06}). The presentation follows the style of Ref. \cite{Silva3}, which contains a more thorough discussion. In addition, in particular, Refs. \cite{Silva, Serbeto, Silva4} were also followed. 

In the semiclassical approximation, the interaction Lagrangian for a neutrino in an electron background reads
\begin{equation}
{\cal L}_{\rm int} = - \frac{G_F}{\sqrt{2}}\left(n_e - \frac{{\bf J}_{e}\cdot{\bf v}_\nu}{c^2}\right)(C_V + 1) \,,
\end{equation}
where ${\bf J}_e = n_{e}{\bf u}_e$, ${\bf v}_\nu$ is the neutrino velocity and $C_V = 1/2 + 2\,\sin^{2}\theta_W$ is the vector-current coupling constant, where $\theta_W$ is the Weinberg mixing angle, with $\sin\theta_W \simeq 1/2$. Therefore, $C_V \simeq 1$. The semiclassical approximation is satisfactory as long as the neutrino de Broglie wavelength $\lambda_\nu = 2\pi\hbar/p_v$ ($p_\nu$ is the neutrino momentum) is much shorter than the typical oscillation length scales. This assumption is expected to be safely true for ultra-relativistic neutrinos. 

The full Lagrangian for a neutrino includes the free Lagrangian ${\cal L}_0$ for a spinless massive particle, so that 
\begin{equation}
{\cal L} = {\cal L}_0 + {\cal L}_{\rm int} = - m_\nu c^2 \sqrt{1 -  v_{\nu}^2/c^2} - \sqrt{2}\, G_F \left(n_e - \frac{{\bf J}_{e}\cdot{\bf v}_\nu}{c^2}\right) \,,
\end{equation}
where $m_\nu$ is the neutrino mass. 

The Hamiltonian formulation is found \cite{Silva3} to be more straightforward to build a theory of the electrons and neutrinos coupling. Therefore, we compute the neutrino canonical momentum, 
\begin{equation}
\label{p}
{\bf P}_\nu = \frac{\partial {\cal L}}{\partial{\bf v}_\nu} = {\bf p}_\nu + \sqrt{2}\,\frac{G_F}{c^2}{\bf J}_e \,, \quad {\bf p}_\nu = \frac{m_\nu {\bf v}_\nu}{\sqrt{1 -  v_{\nu}^2/c^2}} \,,
\end{equation}
and the Hamiltonian, 
\begin{equation}
{\cal H} = {\bf P}_{\nu}\cdot{\bf v}_\nu - {\cal L} = \sqrt{\left({\bf P}_{\nu} c - \sqrt{2}\,\frac{G_F}{c}{\bf J}_e\right)^2 + m_{\nu}^2 c^4} + V_{\rm eff} \,,
\end{equation}
where $V_{\rm eff} = \sqrt{2}\, G_F n_e$ is an effective repulsive potential between neutrinos and the plasma electrons. 

In component-wise form, the canonical momentum equation is
\begin{eqnarray}
\frac{d P_{\nu\,i}}{dt} &=& - \frac{\partial{\cal H}}{\partial r_i} = - \frac{\partial V_{\rm eff}}{\partial r_i} + \sqrt{2}\,G_F 
\sum_{j=1}^{3}\frac{(P_{\nu\,j} - \sqrt{2} G_F J_{e\,j}/c^2)}{\sqrt{({\bf P}_{\nu} c - \sqrt{2}\,G_F\,{\bf J}_{e}/c)^2 + m_{\nu}^2 c^4}}\,\frac{\partial J_{e\,j}}{\partial r_i} \nonumber \\
&=& 
- \frac{\partial V_{\rm eff}}{\partial r_i} + \sqrt{2}\,\frac{G_F}{c^2} \sum_{j=1}^{3}v_{\nu\,j}\frac{\partial J_{e\,j}}{\partial r_i} \,, \quad i = 1, 2, 3, \quad {\bf v}_\nu = \frac{d{\bf r}}{dt} \,. \label{pp}
\end{eqnarray}

Using Eqs. (\ref{p}) and (\ref{pp}) the equation for the mechanical momentum is found to be
\begin{equation}
\frac{dp_{\nu\,i}}{dt} = - \sqrt{2} G_F \left[\frac{\partial n_e}{\partial r_i} + \frac{1}{c^2}\frac{\partial J_{ei}}{\partial t} 
+ \sum_{j=1}^{3}\frac{v_{\nu\,j}}{c^2}\left(\frac{\partial J_{e\,i}}{\partial r_j} - \frac{\partial J_{e\,j}}{\partial r_i}\right)\right] \,,
\end{equation}
which, after a rearrangement, accounts for the neutrino momentum transport equation (\ref{eq06}). Although the above derivation applies to a single neutrino, the fluid description follows in the spirit of the wave packet formalism \cite{Kayser} and the replacement ${\bf v}_\nu \rightarrow {\bf u}_\nu$, the neutrino fluid velocity. An alternative approach for the same problem starts from the kinetic theory for neutrinos in an ionized medium \cite{Silva3}, which is justified by Finite Temperature Quantum Field Theory methods \cite{Semikoz}.  

So far, only the effect of the plasma electrons on neutrinos has been studied. It is found that the neutrino bunching due to the interaction with the collective modes causes a neutrino fluid pressure gradient, and hence gives rise to a ponderomotive force on the electron fluid. 
We refer the reader to Eq. (20) of Ref. \cite{Silva3} and the associated reasoning around it, for the detailed derivation of the neutrino ponderomotive force ${\bf F}_\nu$ in our Eqs. (\ref{eq02}) and (\ref{eq07}).

\end{document}